# Temperature of the Source Plasma for Impulsive Solar Energetic Particles

**Donald V. Reames[1] • Edward W. Cliver[2] • and Stephen W. Kahler[3]**

[1]Institute for Physical Science and Technology, University of Maryland, College Park, MD 20742-2431 USA, email: dvreames@umd.edu

[2]Space Vehicles Directorate, Air Force Research Laboratory, Sunspot, NM 88349, USA, email: ecliver@nso.edu

[3]Air Force Research Laboratory, Space Vehicles Directorate, 3550 Aberdeen Avenue, Kirtland AFB, NM 87117, USA, email: stephen.kahler@kirtland.af.mil

**Abstract** The steep power-law dependence of element abundance enhancements on the mass-to-charge ratios [$A/Q$] of the ions in impulsive solar energetic-particle (SEP) events causes these enhancements to reflect the temperature-dependent pattern of $Q$ of the ions in the source plasma. We searched for SEP events from coronal plasma that is hotter or cooler than the limited region of 2.5 – 3.2 MK previously found to dominate 111 impulsive SEP events. Fifteen new events were found, four (three) originated in 2-MK (4-MK) plasma, but none from outside this temperature range. Although the impulsive SEP events are strongly associated with flares, this result indicates that these ions are not accelerated from flare-heated plasma, which can often exceed 10 MK. Evidently the ions of 2 – 20 MeV amu$^{-1}$ that we observe in space are accelerated from active-region plasma on open magnetic-field lines near the flare, but not from the closed loops of the flare. The power-law dependence of the abundance enhancements on $A/Q$ of the ions is expected from theoretical models of acceleration from regions of magnetic reconnection.

# 1. Introduction

The relative abundances of the chemical elements have provided great insight into the physics of the sources of a variety of energetic-ion populations seen throughout the heliosphere (*e.g.* Reames 1999. However, it is only recently that we have realized the full consequences of elemental abundances of impulsive solar energetic particles (SEPs) that depend upon the temperatures of their source plasma in the solar corona (Reames, Cliver, and Kahler 2014a, b, hereinafter Articles 1 and 2). Since the ionization state of an element [$Q$] depends upon temperature, and since abundance enhancements in impulsive-flare-associated SEPs have a steep power-law dependence upon the ion mass-to-charge ratio [$A/Q$] the temperature-dependent pattern of $A/Q$ is reflected in the pattern of observed enhancements and can be studied on an event-by-event basis for solar-flare-associated impulsive SEP events (Article 2).

Since the pioneering review of Meyer (1985), it has been clear that the abundances of elements in large "gradual" SEP events are closely related to the corresponding abundances of elements in the solar corona. SEPs in gradual events are accelerated, in proportion to the ambient coronal "seed population," by shock waves, driven out from the Sun by fast, wide coronal mass ejections (CMEs). These abundances in gradual SEP events contrast sharply with the spectacular abundance enhancements of the smaller, more numerous, "impulsive" SEP events that have 1000-fold enhancements of both $^3$He/$^4$He and heavy elements, *i.e.* ($Z \geq 50$)/O, resulting from acceleration in solar flares and jets (for a recent review of gradual and impulsive SEP events see Reames, 2013). The distinction of gradual and impulsive SEP events involves a wide variety of evidence including review articles ( Meyer, 1985; Reames, 1990, 1995b, 1999, 2002, 2013; Gosling, 1993; Lee, 1997; Tylka, 2001; Mason 2007), abundances (Reames 1995a; Slocum *et al.*, 2003; Cohen *et al.*, 2007; Mewaldt *et al.* 2007), isotopes (Temerin and Roth, 1992; Roth and Temerin, 1997; Liu, Petrosian, and Mason 2006; Mason 2007; Leske *et al.*, 2007) electrons (Cliver and Ling, 2007, 2009; Wang *et al.*, 2012), impulsive SEP events (Reames, Meyer, and von Rosenvinge, 1994), reacceleration (Mason Mazur, and Dwyer, 1999; Desai *et al.*, 2003; Tylka, *et al.*, 2005; Tylka and Lee, 2006), shock theory (Lee, 2005; Ng and Reames, 2008; Sandroos and Vainio, 2009), shock SEP observations (Kahler 2001; Desai *et al.*, 2004, 2006), relationship to CMEs and shocks (Kahler *et al.* 1984; Kahler, 1992, 1994; Gopalswamy *et al.*, 2002; Cliver, Kahler, and Reames, 2004; Rouillard *et al.*, 2011, 2012) and particle transport (Ng, Reames, and Tylka, 2003).

The relatively recent expansion of measurements to elements throughout the rest of the periodic table above Fe (Reames, 2000; Mason *et al.*, 2004; Reames and Ng, 2004; Mason, 2007; Articles 1 and 2) has contributed significantly to the discrimination and characterization of impulsive SEP events.

The characteristic enhancement in $^3$He/$^4$He, a historic attribute of impulsive SEP events (see Reames 2013), apparently arises from resonant wave particle interactions in the flare plasma, specifically from electromagnetic ion-cyclotron waves produced near the gyrofrequency of $^3$He by copious electron beams streaming out from the flare (Temerin and Roth, 1992; Roth and Temerin, 1997; see also Liu, Petrosian, and Mason 2006). However, the smooth rise of element abundance enhancements with increasing $A/Q$ is not fit well by the resonance model. The fact that the magnitude of the heavy-element enhancements are uncorrelated with those of enhancements in $^3$He/$^4$He (*e.g.* Mason *et al.*, 1986; Reames, Meyer, and von Rosenvinge, 1994; Article 2) suggests the operation of two physical mechanisms. Recently, the heavy-element enhancements, relative to coronal abundances, and their strong power-law dependence on $A/Q$ of the ions in impulsive SEP events, have been linked theoretically to the physics of magnetic-reconnection regions from which the ions escape (Drake *et al.*, 2009, 2010; Knizhnik, Swizdak, and Drake 2011; Drake and Swizdak, 2014). This model is based upon particle-in-cell (PIC) calculations. Ions must first be pre-accelerated to the Alfvén speed in the exhaust from a magnetic reconnection region where the minimum ion $A/Q$ depends upon the plasma $\beta$ which decreases with time (Drake and Swisdak, 2014). Ions are then fast enough to undergo Fermi acceleration by scattering back and forth from the ends of magnetic islands of reconnection as they approach each other, attaining power-law spectra (Drake *et al.* 2009, 2010, Drake, Swisdak, and Fermo 2013). The highest rigidity (highest A/Q) ions escape earliest before the islands collapse. It is generally recognized that the $Q$ of ions in impulsive SEP sources differs from that seen at 1 AU because the ions are stripped of additional electrons by passing through small amounts of material in transit through the corona (Kartavykh *et al.* 2002; Klecker *et al.* 2006).

Regarding the solar associations of gradual and impulsive SEP events, Kahler *et al.* (1984) established a clear (96%) association of the *gradual* events with fast, wide CMEs suggesting acceleration at the CME-driven shock wave. However, Kahler, Reames, and Sheeley (2001) showed that some classic *impulsive* SEP events were associated with narrow CMEs and Yashiro *et al.* (2004) found that at least 28 – 39% of impulsive SEPs had associations with CMEs as well as having their previously known association with solar flares and type-III radio bursts. In Article 1, we speculated that all (impulsive) Fe-rich events have associated CMEs. These associations have tied impulsive SEP events to the mechanism described by the theory of jets (Shimojo and Shibata, 2000) where emerging magnetic flux reconnects on open-field lines allowing easy escape of SEPs and of plasma, *i.e.* the CME (Shimojo and Shibata, 2000; Kahler, Reames, and Sheeley, 2001; Reames, 2002; Nitta *et al.*, 2006; Wang, Pick, and Mason, 2006; Rust *et al.* 2008; Moore *et el.*, 2010; Archontis and Hood, 2013). As pointed out in Article 1, these "open-field-line" CMEs are distinct from the CMEs resulting from eruptions of magnetically closed structures that are associated with gradual SEP events.

If the abundance enhancements in impulsive SEP events depend, both observationally and theoretically, on a power of $A/Q$ of the ions, then since $Q$ depends upon the temperature in the source plasma, the pattern of enhancements provides information on the source-plasma temperature. This property was used to determine that the most probable temperature region was near 3 MK for average impulsive events (Article 1) and to find the variations over 111 individual impulsive SEP events (Article 2). Surprisingly, perhaps, nearly all of the events (108/111) showed source-plasma temperatures in the limited range 2.5 – 3.2 MK.

Figure 1 shows the theoretical dependence of $A/Q$ on temperature. An early study (Reames, Meyer, and von Rosenvinge 1994) found enhancements of major ions in three groups: i) He, C, N, and O ii) Ne, Mg, and Si, and iii) Fe. It was suggested that this pattern corresponded to $\approx 3 - 5$ MK.

The references to the data on ionization states that we used, listed in the caption of Figure 1, were chosen for comparability with previous work. More recent information, *e.g.* Mazzotta *et al.* (1998) differs imperceptibly from these historic references in our region of interest.

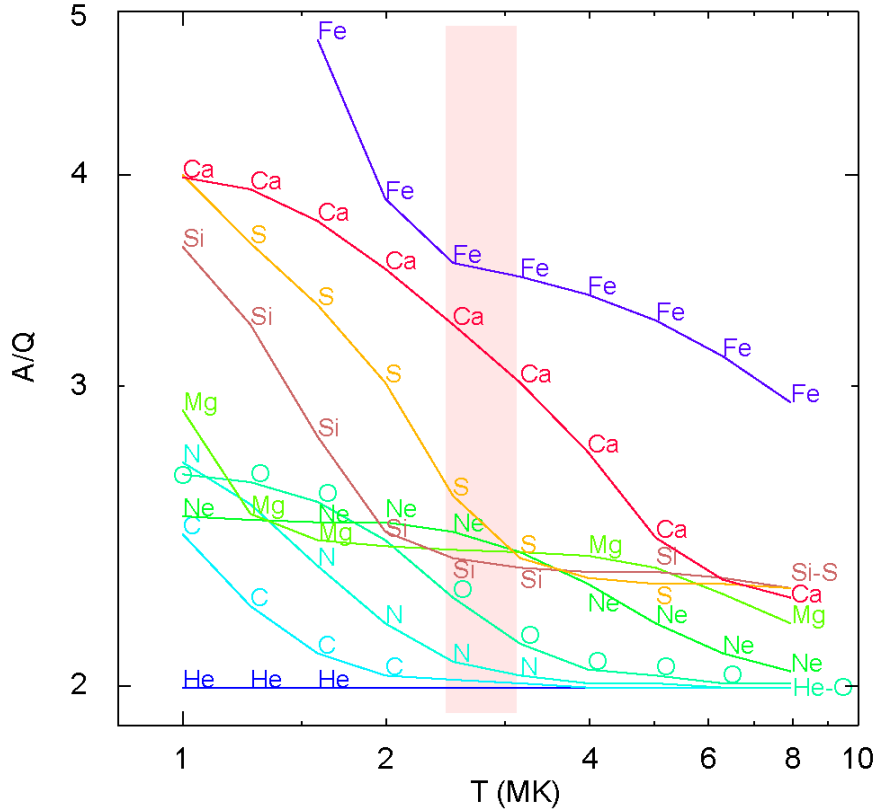

**Figure 1.** *A/Q* is plotted as a function of the theoretical equilibrium temperature for several elements. Elements below Fe are from Arnaud and Rothenflug (1985), Fe from Arnaud and Raymond (1992) and elements in the high-Z region from Post *et al.* (1977). Elements are named along each curve. The region of the most likely temperature for Fe-rich impulsive SEP events (Articles 1 and 2) is shaded pink.

With improved measurements from the *Wind* spacecraft (Article 1), it became clear that: i) in many events, the enhancement in Ne exceeded that in Mg and in Si, out of Z-order, ii) some "He-poor" events had He/O well below the coronal value arising from enhancements in O but not in He, and iii) power-law fits of enhancements *vs. A/Q* extended smoothly to the region of atomic numbers $76 \leq Z \leq 82$ for temperatures $T \approx 3$ MK. Figure 1 shows that the region where the *A/Q* of Ne exceeds that of Mg and Si and where *A/Q* of O rises above two occurs uniquely near 2.5 MK. Here the power-law fits of enhancements *vs. A/Q* for each event (Article 2) select the shaded temperature region (2.5 − 3.2 MK) in Figure 1.

The impulsive SEP events listed in Article 1 were selected by requiring events to be Fe-rich, specifically, an enhancement (Fe/O)/0.131 > 4, where 0.131 is the assumed coronal value of Fe/O. While the *A/Q* values in Figure 1 suggest that Fe should be enhanced throughout the temperature range, we should be able to identify abundance patterns that would be more typical of coronal temperatures that are hotter or cooler than the 2.5 – 3.2 MK band.

Have we overlooked impulsive SEP events from hotter or cooler source plasma by the selection criteria of Article 1? The present paper seeks to extend the event list of Articles 1 and 2 by searching specifically for impulsive SEP events with abundance patterns appropriate to source temperatures outside the very limited temperature range found for the 111 events studied in Articles 1 and 2.

The SEP observations for this article were measured by the *Low Energy Matrix Telescope* (LEMT: von Rosenvinge *et al.*, 1995) onboard the *Wind* spacecraft which measures elements from He through about Pb in the energy region from about 2 – 20 MeV amu$^{-1}$ with a geometry factor of 51 cm$^2$ sr, identifying and binning the major elements from He to Fe onboard at a rate up to about $10^4$ particles s$^{-1}$. Instrument resolution and aspects of the processing have been shown and described elsewhere (Reames *et al.*, 1997; Reames, Ng, and Berdichevsky, 2001; Reames, 2000; Reames and Ng, 2004). Element resolution in LEMT from C – Fe, as a function of energy, has been shown recently and abundances in gradual events have been analyzed by Reames (2014). Typical resolution of LEMT from He isotopes through Fe was shown by Reames *et al.* (1997) and resolution of elements with $34 < Z < 82$ by Reames (2000). The LEMT response was calibrated with accelerator beams such as C, O, Fe, Ag, and Au before launch (von Rosenvinge *et al.*, 1995).

## 2. SEP Events from Cooler Source Plasma

At temperatures of 2 MK and below, Figure 1 shows that O has joined the elements with only two orbital electrons, such as Ne, while He remains at $A/Q = 2$. Lowering $Q$ for O raises its value of $A/Q$ and, hence, its abundance. Thus, events from this region should all be "He poor." We scanned the LEMT data from the time period from 3 November 1994 to 5 August 2013, the same period studied in Articles 1 and 2, for He-poor SEP events (with enhancement He/O/47 < 0.4) that are sufficiently intense that abundances of the dominant elements (He, C, N, O, Ne, Mg, Si, and Fe) are measurable. This scan located 26 events, of which 15 had already been measured in Articles 1 and 2. Some abundance ratios in these events are shown in Figure 2.

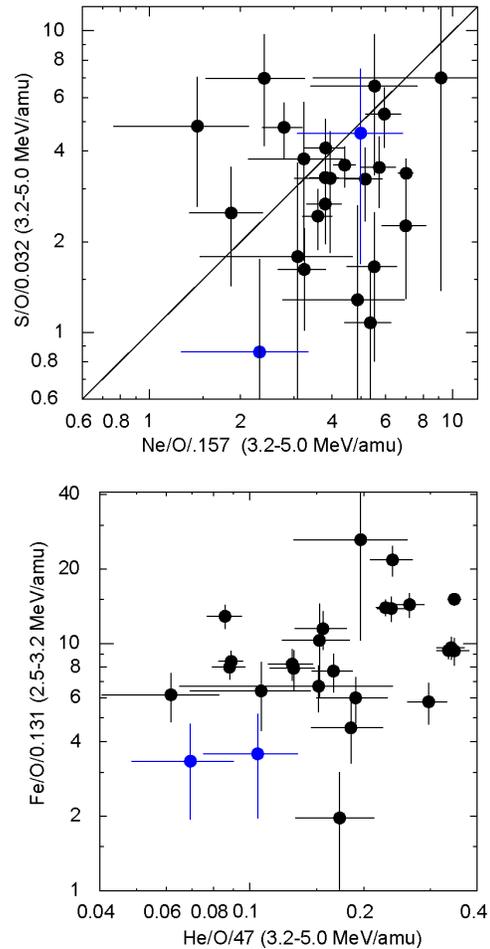

**Figure 2.** Elemental abundance ratios relative to coronal abundances are shown for He-poor events. Enhancements in Fe/O *vs.* He/O are shown in the lower panel and in S/O *vs.* Ne/O in the upper panel. Fe/O is shown for the 2.5 – 3.2 MeV amu$^{-1}$ interval used as a selection criterion in Article 1. Other ratios are in the 3.2 – 5.0 MeV amu$^{-1}$ interval that is available for all species. The lower panel shows that we have accepted three new events with an enhancement of Fe below four. The diagonal line in the upper panel shows that the enhancement in Ne/O still exceeds that in S/O for most events. The locations of two of the three events with low Fe/O are shown in blue; the third event is too small to have any S.

The lower panel in Figure 2 shows that all the events are Fe-rich, but three of the new events are below the criterion for the original list. The upper panel compares S/O with Ne/O. From Figure 1 we see that $A/Q$ for S rises rapidly above that for Ne below 2.5 MK. We know that these new events are all at or below 2.5 MK because they are required to be He-poor events where the A/Q of O exceeds that of He. Since the enhancement of S/O is comparable or below that of Ne/O in most events shown in Figure 2, we expect that most of these new events are *not* actually at temperatures below 2.5 MK.

Figure 3 shows the abundance enhancements *vs.* *A/Q* at 2.0 MK for an event beginning at 2000 UT on 18 March 1999. Note that O has moved above He into the group with Ne and Si. While Fe falls below the fitted line, the elements at higher *Z* suggest that this may not be significant.

**Figure 3**. Element abundance enhancements are plotted *vs.* *A/Q* at 2.0 MK for an impulsive SEP event. The least-squares fit line is shown.

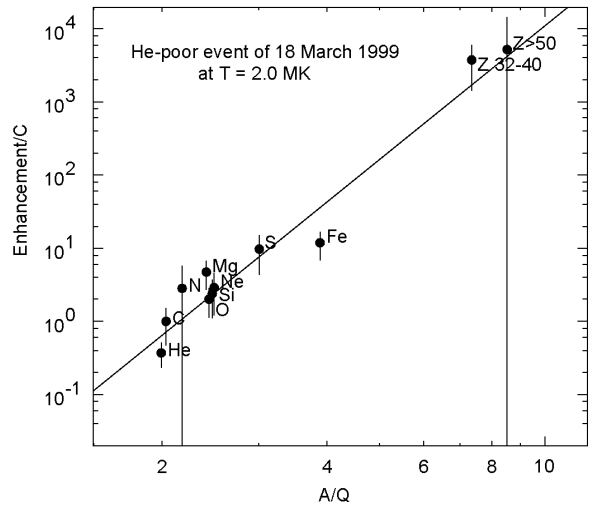

To determine the temperature for each SEP event, we follow the procedure of Article 2. At each temperature we determine the values of *A/Q* and fit enhancement *vs.* *A/Q* as in Figure 3 above, noting the value of $\chi^2$. For each event we then select the temperature and fit that has the lowest value of $\chi^2$. Values of $\chi^2$ *vs.* *T* for the He-poor events are shown in Figure 4.

**Figure 4.** Values of $\chi^2$ *vs.* T are shown for the 26 He-poor events. Events differ in color and symbol. The number of events with minima at each temperature is shown along the bottom of the plot.

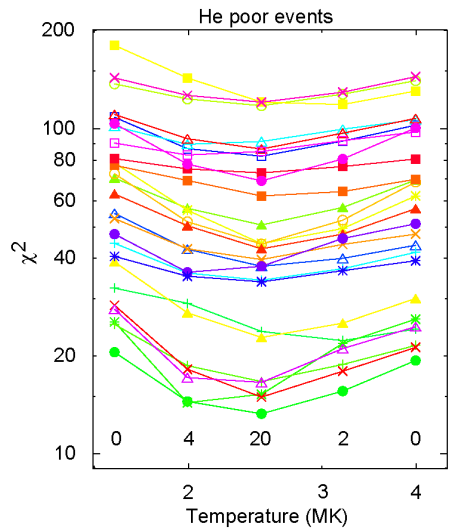

The event shown in Figure 3 is one of the events with $T = 2.0$ MK. As expected from our discussion of S/O *vs.* Ne/O shown in Figure 2, we have been unable to find any events with temperatures below 2.0 MK.

## 3. SEP Events from Hotter Source Plasma

If we examine the region at 4 MK and above in Figure 1, we see that O has $A/Q \approx 2$ so the events should not be He poor (*e.g.* enhancement He/O/47 > 0.5) and the enhancement in Ne should no longer exceed that in Si (Ne/C/0.374 ≤ Si/C/0.360), in fact $A/Q$ for Ne begins to fall toward 2.0. We should also expect the enhancements in S and Si to be comparable. When we scanned the LEMT data from the time period from 3 November 1994 to 5 August 2013 for SEP events with abundance patterns corresponding to these properties in $A/Q$, we found only 15 events, with 11 of these already on the list in Article 1.

Properties of the candidate events from possible "hot" plasma sources are shown in Figure 5. The enhancement of Ne no longer exceeds that of Si and Si and S enhancements are comparable.

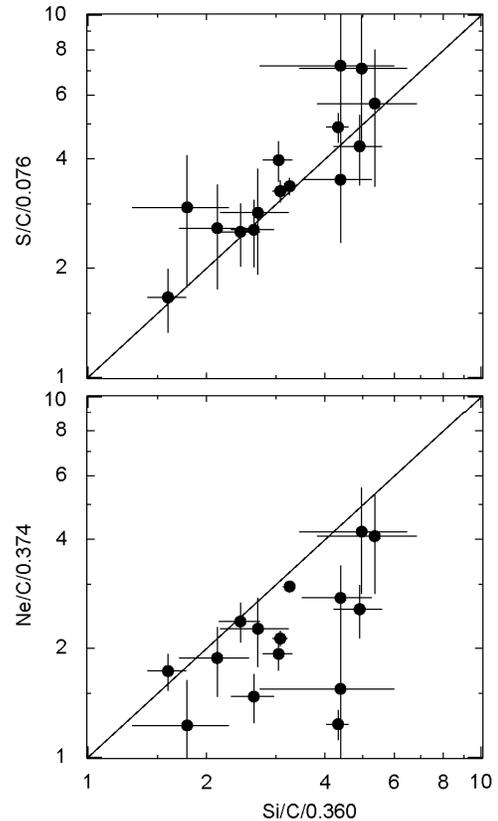

**Figure 5.** Properties are shown of candidate events from hot plasma sources. Relative abundances are shown for Ne/C *vs.* Si/C (lower panel) and S/C *vs.* Si/C (upper panel), each ratio is normalized to the corresponding coronal abundances.

An example of one of the events from this sample with *A/Q* values at 4.0 MK is shown in Figure 6. Note that the elements He, C, N, and O cluster together and that S has joined the cluster of Ne, Mg, and Si.

**Figure 6.** Element abundance enhancements *vs.* *A/Q* at 4.0 MK are shown for a representative event together with a least-squares fit line.

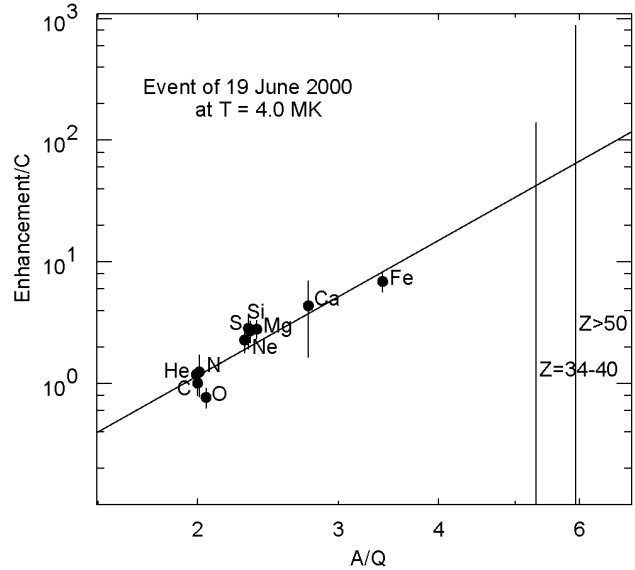

A plot of $\chi^2$ *vs.* *T* for the hot candidate events is shown in Figure 7 with the number of event minima at each temperature shown along the bottom.

**Figure 7.** Values of $\chi^2$ *vs.* T are shown for the 15 "hot" plasma events. Events differ in color and symbol. The number of events with minima at each temperature is shown along the bottom of the plot.

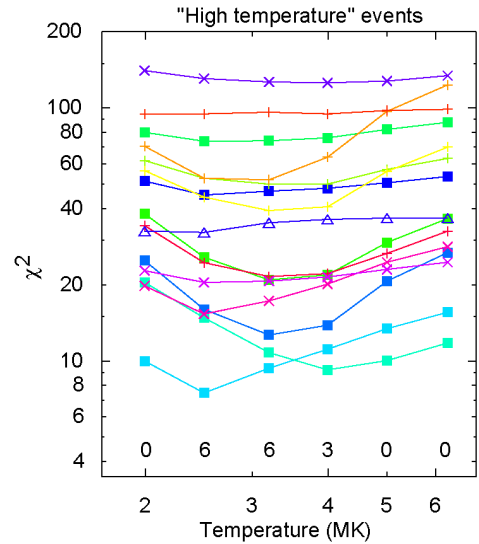

# 4. The Source Temperature Distribution

If we include the 15 new events with the 111 in the original sample, the temperature distribution will be broadened somewhat. Since the weighting of the enhancement of each particle species affects the slope of the fit *vs.* *A/Q*, it can also affect the assigned plasma temperature. As discussed in Article 2, the residuals of the fit show spreads of $15 - 30\%$ at the dominant elements. Weighting with only statistical errors will cause the elements with the most numerous particles, *e.g.* He and Fe, to dominate the fit, even though the spread of He/C is quite large (Article 2). Including a fixed error will increase the relative importance of the elements with less numerous samples, making the contribution of differing species more nearly equal.

In Figure 8 we show the distribution of source-plasma temperatures deduced for the power-law fit with minimum $\chi^2$ for different assumed values of the fixed error in the measured enhancements. Using statistical errors only (shown as 0 %) causes the distribution to be spread to higher temperatures, but 30 % is more representative of the observed spread in He/C, for example. In any case, the temperature range of $2 - 4$ MK dominates.

**Figure 8.** The distribution of source plasma temperature is shown for values of the fixed error (0 %, 20 % and 30 %) included in weighting the enhancements for fits *vs. A/Q*.

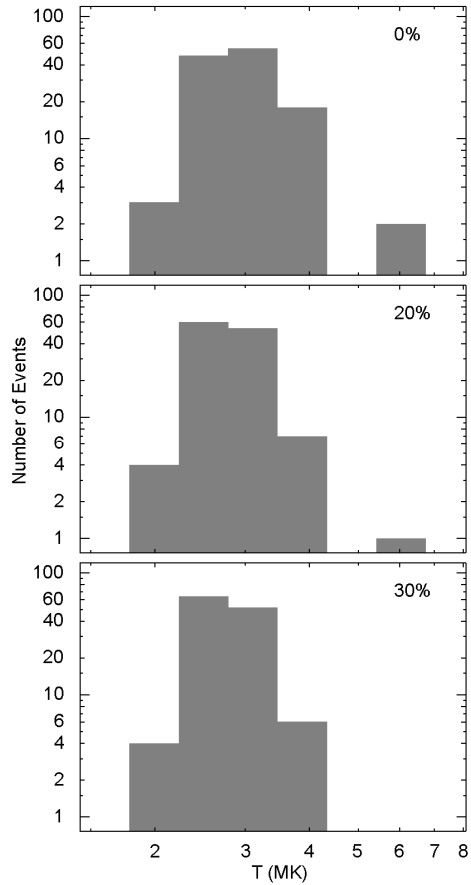

Another reason for considering larger, more realistic values of the fixed error of $\approx 30$ % is their effect on $\chi^2$. When the residuals in least-squares fits balance the errors used for weighting, $\chi^2$ approaches the number of degrees of freedom. In our case, if there are 12 enhancement measurements in an event, we would expect $\chi^2 \approx 10$. For points with small statistical errors, which occur frequently, increas-ing the fixed error by a factor of two, *e.g.* from 15 % to 30 %, can decrease $\chi^2$ by a factor of four. Events with the larger values of $\chi^2 \approx 40$ in Figure 7, for example, can be reduced to $\approx 10$ by such a change. The fixed spread was discussed with respect to residuals in Section 2.3 of Article 2. Some of these variations, in He/C, for example, easily exceed 30 %; they are of unknown origin, but are unlikely to be related to either ionization or to the power of $A/Q$ for this case, since He and C are usually fully ionized with $A/Q = 2$. Nevertheless, even errors of $30 - 40$ % are small compared to the order-of-magnitude variation of enhancements *vs. A/Q*.

## 5. The Distribution of Power-Law Enhancements

The spread in the distribution of the power of $A/Q$ for all of the events is shown in Figure 9. The mean of the distribution, shown in the figure, is $4.47 \pm 0.07$, and the standard deviation of a single event, also shown, is 0.74. This distribution was obtained using 30 % fixed error in the weighting. Using 20 % weighting, the distribution is slightly broader with a mean of $4.44 \pm 0.07$ and a width of 0.81.

**Figure 9.** The distribution of the power of the enhancement *vs. A/Q* is shown for 126 impulsive SEP events. The mean and standard deviation are shown.

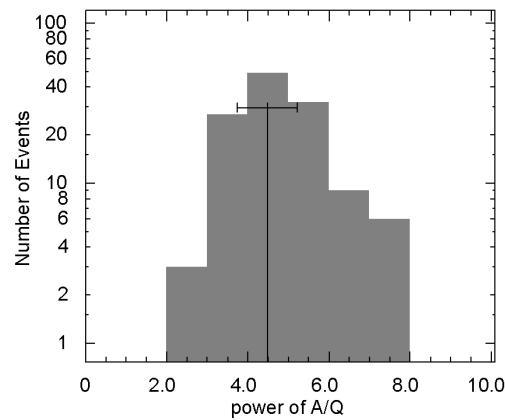

If we compare the distribution of the power of $A/Q$ for different plasma temperatures we find that any evidence of a significant variation is quite weak. The correlation coefficient of the power of the $A/Q$ dependence *vs.* log ($T$) is -0.22.

## 6. Discussion

The most significant result for our study of the source-plasma temperatures leading to observed abundance enhancement patterns in impulsive SEP events is that we have been unable to find evidence for hot $\geq 10$ MK sources that might be expected from X-ray observations of solar flares. Thus, what we did not find may be more significant than what we found. SEPs do not come from the hot plasma in flares. Either the SEPs are accelerated very early or they are accelerated on open magnetic-field lines near the flare where heating is minimal, or both. Open field regions near a flare appear dark, indicating minimal heating (see, *e.g.*, Figure 3 of Rust *et al.* 2008). High flare temperatures are likely to be a signature of re-connection on closed magnetic loops that trap energy to produce heating. On open-field lines the energy is carried away with minimal heating. The temperatures that we have deduced for the SEPs are near the 3 – 4 MK found for active region core temperatures (Warren, Winebarger, and Brooks 2012; Del Zanna and Mason 2014).

We should also note that there is γ-ray evidence (Murphy *et al.* 1991) that the broad (Doppler-shifted) γ-ray lines representing the accelerated "beam" in flares have enhanced Ne/O, just as many of our SEP events do. This enhancement indicates a temperature of $\approx 3$ MK. If these ions are accelerated in flare loops, then they must be accelerated early, before significant heating can occur. Alternatively, they are derived from the same population that we eventually see in space.

SEP ionization states can be measured directly in space. In early measurements at $0.3 – 2$ MeV amu[-1] (Luhn *et al.* 1985), the ionization states were found to indicate a temperature of $2 – 7$ MK, but later study (Luhn *et al.* 1987) found temperatures for [3]He-rich events of $\approx 10$ MK, largely because $Q_{Fe} \approx 20$ was found. At energies >15 MeV amu[-1], Leske *et al.* (2001) used geomagnetic filtering to find that Fe-rich events had $Q_{Fe}$ ~20 while events with low Fe/O had $Q_{Fe} \approx$ 15. More recently, DiFabio *et al.* (2008) found energy dependence where $Q_{Fe}$ increased from $\approx 15$ to $\approx 20$ with ion velocity from $\approx 0.1$ to $\approx 0.5$ MeV amu[-1]. There are now models of the transport of ions to 1 AU that include the process of stripping of ions in transit through the corona from a source at $1 – 10$ MK and produce $Q_{Fe}$ to values near $\approx 20$ with the observed energy dependence (Kartavykh *et al.* 2002; Klecker *et al.* 2006). These models seem quite compatible with our results.

Ions can attain an equilibrium charge state that increases with their velocity if they pass through very small amounts of matter [$\approx$ μg cm[-2]] after acceleration. The ions will be stripped or will pick up electrons as needed to attain this equilibrium (*e.g.* Klecker *et al.* 2001). Thus the energy dependence of $Q_{Fe}$ in im-pulsive SEP events strongly suggests that the ionic charge states found in space have little to do with those that exist during acceleration. Energetic Fe ions from gradual events, accelerated from altitudes of two to three solar radii, show no evidence of stripping, but those from impulsive events, perhaps accelerated low in the corona, do.

Some years ago, Mullan and Waldron (1986) proposed X-ray photo-ionization as a mechanism to produce the observed SEP ionization states *in lieu* of thermal ionization. While this mechanism was used to explain the ionization states observed by Luhn *et al.* (1985), which are now understood to be modified by transport away from the source, the idea can produce a pattern of ionization such as the inverse-*Z* ordering of *A/Q* for Ne, Mg, and Si. This mechanism would seem to lend itself well to the association of SEP events with solar flares. However, one would expect a wide range of ionization patterns from the 1000-fold range of flare X-ray intensities, and, perhaps, even varying SEP source distances from the flaring plasma. Why would we get such similar abundance patterns, such as Ne enhancements, in nearly all impulsive SEP events? Would we not find cases where Ne, Mg, or Si is fully ionized like O (but Fe is not)?

Finally, we should note that the SEP events on the original list in Article 1 were chosen to have high values of Fe/O. While most of these events are surely impulsive SEP events, it is possible that some result from shock acceleration of impulsive suprathermal ions from a previous impulsive SEP event (Tylka *et al.* 2005; Tylka and Lee 2006). These would probably be the longer-duration events with fast, wide CMEs. From the CME associations we estimate that ≤10% of the events could involve reaccelerated impulsive material. However, it is likely that the temperature derived from the abundances in these events still represents a valid source plasma temperature in the original impulsive event. Hence, nearly all (≥ 98 %) of the impulsive SEP events that we can find and measure have source plasma temperatures in the range 2 – 4 MK.

## Disclosure of Potential Conflicts of Interest

The authors declare that they have no conflicts of interest.

**Acknowledgments**: S. Kahler was funded by AFOSR Task 15RV-COR167.